\documentclass[a4paper, 11pt]{article}

\usepackage{a4wide}

\usepackage[english]{babel}
\usepackage[utf8]{inputenc}
\usepackage[T1]{fontenc}

\usepackage{amsthm}

\usepackage[ttscale=.875]{libertine}
\usepackage[libertine]{newtxmath}
\usepackage[scaled=1]{inconsolata}

\usepackage{graphicx}

\theoremstyle{remark}

\usepackage{booktabs}

\usepackage[colorlinks=false, hidelinks]{hyperref}

\usepackage{siunitx}
\usepackage{ragged2e}
\newcolumntype{R}[1]{>{\RaggedRight\arraybackslash}p{#1}}

\title{Cryptographic Keywords in NVD: Statistics and Visualization}
\author{
  Martin Stanek \\[2ex]
  Department of Computer Science \\
  Faculty of Mathematics, Physics and Informatics \\
  Comenius University \\
  \textsl{martin.stanek@fmph.uniba.sk}
}

\date{}

\begin{document}
\maketitle

\begin{abstract}
A preliminary attempt to use cryptographic keywords and analyze vulnerabilities published in the National Vulnerability Database is presented. Basic statistics and visualizations are included. \\[2ex]
\end{abstract}

\section{Introduction}

Cryptography plays an important role in securing information systems, networks, and other IT services. Similarly to other security controls, the cryptographic controls can fail when ignored or improperly designed, implemented, or operated. The OWASP lists \textsl{Cryptographic Failures} as the second most critical security risk to web applications in their 2021 edition of OWASP Top 10 \cite{owasp}.

Even though cryptographic vulnerabilities are less frequent than SQL Injections, Cross-site scripting, or Out-of-bound Writes, they still can be potentially exploited with severe consequences. Cryptographic failures were analyzed in general setting using published vulnerabilities \cite{L14}, as well as empirically by analyzing misuses of cryptographic APIs in existing code bases, such as applications written in Java \cite{E13}, C/C++ \cite{Z19}, or Python \cite{W21}.

The National Vulnerability Database (NVD) \cite{nvd}, managed by the National Institute of Standards and Technology (NIST), is the most respected repository of software vulnerabilities, uniquely identified by CVE\footnote{Common Vulnerabilities and Exposures (\url{https://cve.mitre.org/})} identifiers. Nowadays, it publishes a vast number of vulnerabilities, e.g., more than \num{8500} CVEs in the third quarter of 2024. The vulnerabilities are classified according to their underlying weakness, and one or multiple CWE\footnote{Common Weakness Enumeration (\url{https://cwe.mitre.org/}). Note that the NVD uses a selected subset of CWEs.} identifiers are assigned to each CVE.

There are CWEs that are specifically linked to cryptographic issues, such as lack of cryptography in CWE-312 (Cleartext Storage of Sensitive Information) or CWE-319 (Cleartext Transmission of Sensitive Information), implementation issues in CWE-295 (Improper Certificate Validation), design problems in CWE-321 (Use of Hard-coded Cryptographic Key) etc. However, a significant number of CVEs are assigned special CWE identifiers \textsl{NVD-CWE-noinfo} for vulnerabilities with insufficient information to classify them, and \textsl{NVD-CWE-Other} for weaknesses outside of the NVD selection. For example, 15.1\% of CVEs published in the third quarter of 2024 have one of these two assignments.

Keyword analysis of CVE descriptions can provide an interesting and potentially useful insight into published CVEs. It is a supplementary analysis that does not rely on classifications already available in the NVD.

\subsection*{Keywords}

The selection of keywords is subjective, and different sets can be used for other analyses. However, the chosen keywords aim to cover basic cryptographic terms, constructions, algorithms, protocols, and attacks. They are specifically selected to indicate with high probability a cryptographic issue or at least the use of cryptographic constructions. Table \ref{tab-keywords} shows the keywords used in this paper.

\begin{table}[ht]
\centering
\begin{tabular}{llll}
\toprule
encrypt & plaintext & salt & elliptic \\
decrypt & cleartext & password & cryptanalysis \\
cipher & public key & TLS & cryptographic \\
hash & private key & RSA & side channel \\
symmetric & certificate & AES & man in the middle \\
asymmetric & key exchange & ECDSA & replay attack \\
signing & Diffie & ECDH & brute force \\
signature & random & HMAC & \\
\bottomrule
\end{tabular}
\caption{Cryptographic keywords}
\label{tab-keywords}
\end{table}

\section{The dataset and results}

The dataset consists of CVEs published between 2023-01-01 and 2024-09-30 (i.e. the entire year 2023 and three quarters of 2024), downloaded using NVD's public API. The CVEs with status \textsl{Rejected} were filtered out. This results in \num{57773} CVEs in non-rejected state, with an average CVSS\footnote{Common Vulnerability Scoring System, version 3.1} base severity score \num{7.10}.

Matching a keyword in the description is performed as a simple substring search, with some customization: case non-sensitive search, hyphens replaced by spaces, special treatment of some keywords\footnote{For example, RSA was treated separately to avoid matching with ``path trave\textbf{rsa}l'' issues.}. Substring search was chosen to match \textsl{encrypt} with \textsl{encrypt}, \textsl{encryption}, \textsl{encrypted}, etc.

There are \num{3914} CVEs (6.8\%) containing at least one keyword in their description. The average base severity score of these CVEs is 7.18, slightly higher than the score for the entire dataset.

\subsection{Keywords statistics}

Table \ref{tab-keystats} on page \pageref{tab-keystats} displays for each keyword:
\begin{itemize}
\item CVE -- the count of CVE's in which the description includes the keyword, referred to as related CVEs. Unsurprisingly, \textsl{password} is the most frequently mentioned term. Interestingly, some keywords with narrow semantics like \textsl{man in the middle} are mentioned more frequently than some generic terms like \textsl{cipher} or \textsl{public key}.
\item Unique CWE -- the number of unique CWEs identified in the related CVEs. The values are relatively high for the majority of keywords. It indicates that even specific issues (e.g., \textsl{man in the middle} attacks) cannot be analyzed just by focusing on a few CWEs.
\item Avg Score -- average base severity score of related CVEs. Red color indicates entries with scores higher than the dataset's average score.
\end{itemize}

\newcommand{\rt}[1]{\textcolor{red}{#1}}

\begin{table}[ht]
\centering
\begin{tabular}{lrrrr}
\toprule
Keyword & CVE & Unique CWE & Avg Score \\
\midrule
password & 1610 & 164 & \rt{7.54} \\
encrypt & 530 & 127 & 6.73 \\
hash & 381 & 100 & 6.75 \\
certificate & 348 & 92 & 6.97 \\
TLS & 290 & 100 & 7.02 \\
cryptographic & 241 & 72 & \rt{7.36} \\
signature & 215 & 79 & \rt{7.14} \\
decrypt & 206 & 69 & 6.93 \\
random & 187 & 76 & \rt{7.33} \\
man in the middle & 185 & 59 & 6.29 \\
brute force & 179 & 48 & \rt{7.73} \\
cleartext & 164 & 33 & 6.50 \\
plaintext & 146 & 51 & 6.69 \\
side channel & 92 & 13 & 6.19 \\
signing & 88 & 54 & \rt{7.22} \\
cipher & 79 & 39 & 6.51 \\
private key & 79 & 39 & 6.63 \\
AES & 65 & 39 & 6.28 \\
public key & 59 & 38 & 7.05 \\
RSA & 56 & 26 & 6.64 \\
HMAC & 33 & 30 & \rt{7.62} \\
salt & 28 & 24 & 6.89 \\
ECDSA & 22 & 21 & \rt{7.29} \\
replay attack & 22 & 9 & \rt{7.40} \\
symmetric & 21 & 20 & 6.65 \\
key exchange & 17 & 12 & 7.04 \\
ECDH & 12 & 9 & 6.21 \\
elliptic & 12 & 11 & \rt{7.83} \\
asymmetric & 6 & 8 & 7.07 \\
Diffie & 5 & 9 & 6.93 \\
cryptanalysis & 1 & 2 & 6.80 \\
\bottomrule
\end{tabular}
\caption{Keywords statistics}
\label{tab-keystats}
\end{table}

\subsection*{CWE statistics}

Table \ref{tab-cwestats} on page \pageref{tab-cwestats} shows top 25 CWEs sorted by a keyword count. A CVE $c$ is assigned to one or multiple CWEs $w(c) = \{w_1, \ldots, w_{c_n}\}$, and its description contains a set of keywords $k(c) = \{k_1, \ldots, k_{c_m}\}$. The keyword count for a weakness $\mathbf w$ is defined as $\text{kc}({\mathbf w}) = \sum_{c;\; \mathbf w\in w(c)} |k(c)|$, with the understanding that multiple occurrences of a keyword in the same description are counted as a single occurrence. Note that there is only a weak correlation between keyword count and individual CWE frequency. An alternative method, which is not used in this analysis, would involve scaling the keyword count by the number of CVEs: $\text{kc}({\mathbf w}) / |\{c; {\mathbf w}\in w(c)|$. Table \ref{tab-cwestats} includes the following columns:

\begin{itemize}
\item Keywords -- the keyword count associated with a particular CWE.
\item Avg Score -- the average base severity score for the CVEs whose descriptions contain at least one keyword and are associated with the CWE. Entries in red color indicate scores higher than the the score in the last column.
\item Dataset -- the average base severity score for all CVEs assigned to the CWE, calculated from the full dataset.
\end{itemize}

\begin{table}[ht]
\centering
\begin{tabular}{lR{6cm}rrr}
\toprule
CWE & Title & Keywords & Avg Score & Dataset \\
\midrule
NVD-CWE-noinfo &  & 413 & \rt{6.86} & 6.84 \\
CWE-295 & Improper Certificate Validation & 298 & 6.97 & 7.02 \\
CWE-798 & Use of Hard-coded Credentials & 235 & 8.43 & 8.46 \\
CWE-203 & Observable Discrepancy & 225 & \rt{6.11} & 6.00 \\
CWE-200 & Exposure of Sensitive Information to an Unauthorized Actor & 224 & \rt{6.56} & 6.36 \\
CWE-312 & Cleartext Storage of Sensitive Information & 212 & \rt{6.27} & 6.25 \\
CWE-327 & Use of a Broken or Risky Cryptographic Algorithm & 180 & \rt{7.33} & 7.28 \\
CWE-287 & Improper Authentication & 156 & 7.62 & 7.89 \\
CWE-89 & Improper Neutralization of Special Elements used in an SQL Command ('SQL Injection') & 156 & \rt{8.93} & 8.92 \\
CWE-319 & Cleartext Transmission of Sensitive Information & 147 & 6.88 & 6.94 \\
CWE-522 & Insufficiently Protected Credentials & 146 & \rt{6.86} & 6.78 \\
CWE-347 & Improper Verification of Cryptographic Signature & 145 & 7.28 & 7.32 \\
CWE-326 & Inadequate Encryption Strength & 143 & \rt{7.01} & 6.82 \\
NVD-CWE-Other &  & 114 & \rt{6.95} & 6.88 \\
CWE-307 & Improper Restriction of Excessive Authentication Attempts & 110 & \rt{8.14} & 8.01 \\
CWE-79 & Improper Neutralization of Input During Web Page Generation ('Cross-site Scripting') & 97 & \rt{5.91} & 5.66 \\
CWE-532 & Insertion of Sensitive Information into Log File & 96 & \rt{6.13} & 6.00 \\
CWE-311 & Missing Encryption of Sensitive Data & 80 & 6.02 & 6.18 \\
CWE-321 & Use of Hard-coded Cryptographic Key & 72 & \rt{8.13} & 8.06 \\
CWE-330 & Use of Insufficiently Random Values & 72 & 7.22 & 7.32 \\
CWE-121 & Stack-based Buffer Overflow & 71 & 8.47 & 8.68 \\
CWE-521 & Weak Password Requirements & 70 & \rt{8.01} & 7.97 \\
CWE-284 & Improper Access Control & 67 & \rt{8.04} & 6.95 \\
CWE-787 & Out-of-bounds Write & 65 & 7.87 & 7.95 \\
CWE-256 & Plaintext Storage of a Password & 62 & \rt{7.09} & 6.58 \\
\bottomrule
\end{tabular}
\caption{CWE statistics (Top 25 CWEs according to the keyword count)}
\label{tab-cwestats}
\end{table}

\subsection*{Heatmaps}

\begin{figure}[ht]
\centering
\includegraphics[scale=0.62, angle=-90]{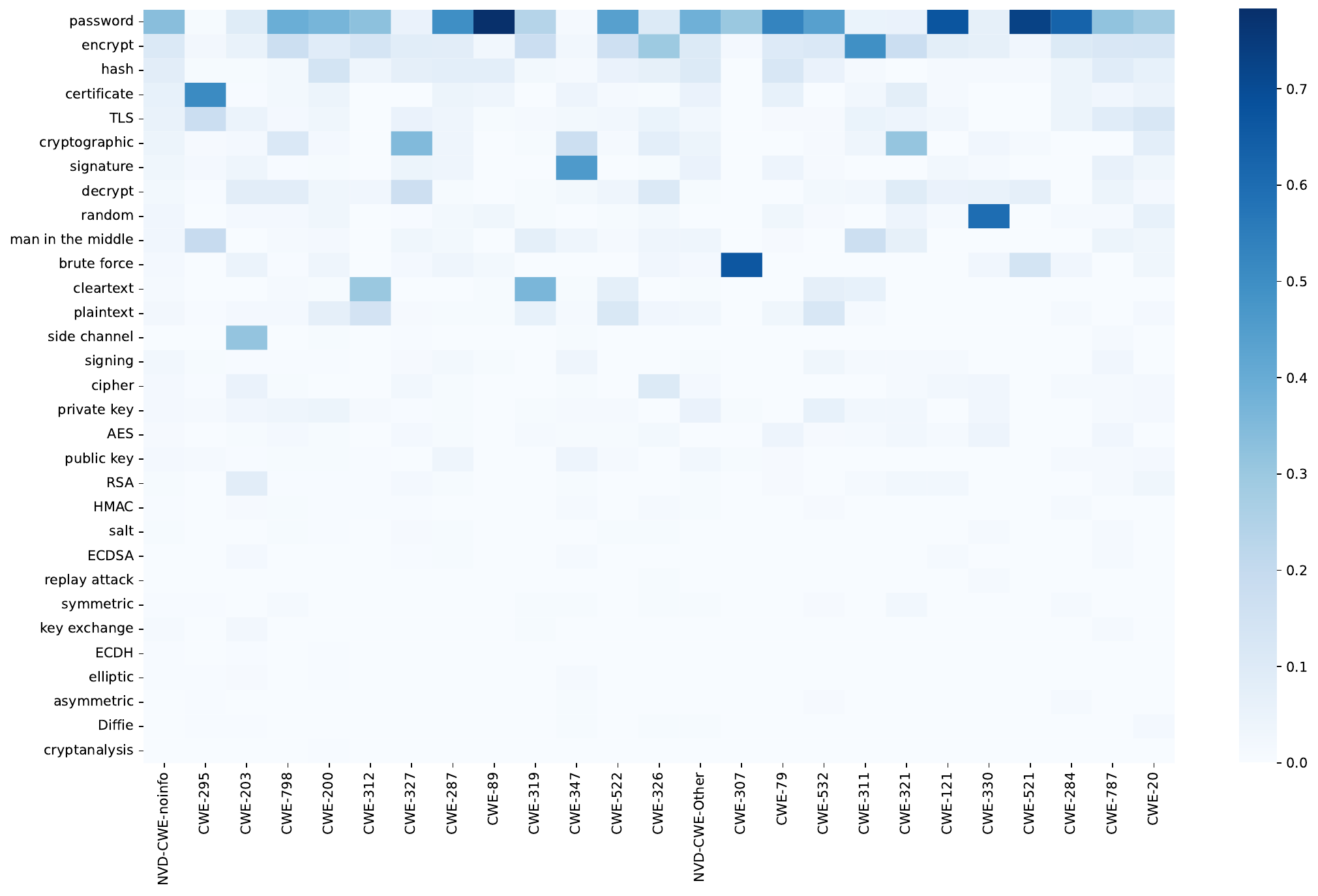}
\caption{Correlation of keywords and CWEs, based on vulnerabilities with descriptions containing at least one keyword, normalized for each CWE.}
\label{fig-keywords}
\end{figure}

The relation between individual keywords and CWEs can be visualized through a heatmap. Figure \ref{fig-keywords} on page \pageref{fig-keywords} illustrates the prevalence of each keyword in the descriptions of CVEs assigned to specific CWEs. Again, only the top 25 CWEs are analyzed. As expected, \textsl{password} is a predominant keyword across most CWEs. The exceptions are usually particular CWEs correlated with obvious keywords, for example:
\begin{itemize}
\item CWE-295 Improper Certificate Validation -- \textsl{certificate};
\item CWE-203 Observable Discrepancy -- \textsl{side channel};
\item CWE-319 Cleartext Transmission of Sensitive Information -- \textsl{cleartext};
\item CWE-307 Improper Restriction of Excessive Authentication Attempts -- \textsl{brute force}; 
\item CWE-330 Use of Insufficiently Random Values -- \textsl{random}, etc.
\end{itemize}

Examining the heatmap further, we can find the most relevant CWEs to a specific keyword. As an example \textsl{man in the middle} seems to be related mostly to CWE-295, CWE-311, CWE-321, and CWE-319.

\medskip

\begin{figure}[ht]
\centering
\includegraphics[scale=0.62, angle=-90]{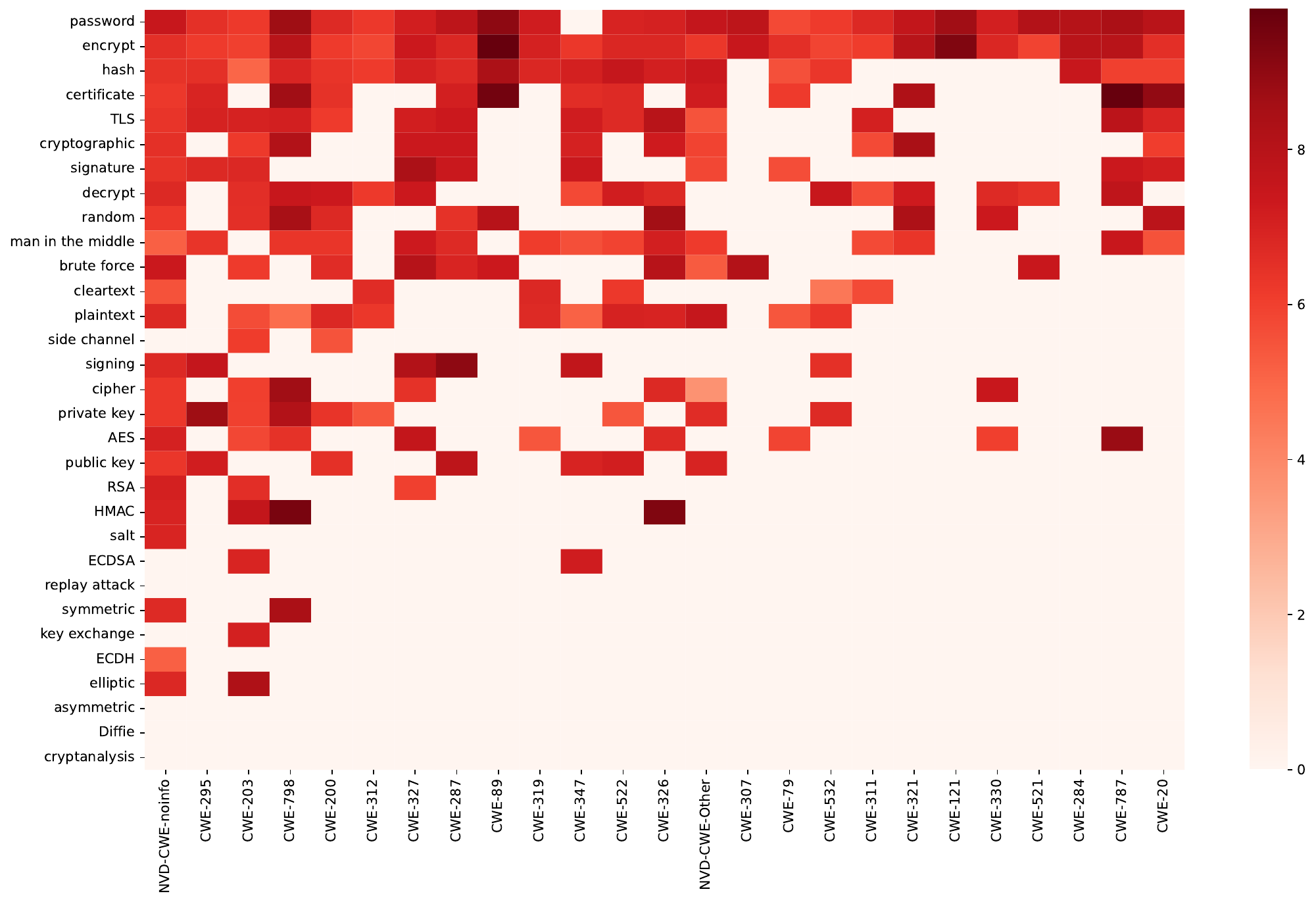}
\caption{Average CVSS base severity score of vulnerabilities assigned to CWE with description containing given keyword.}
\label{fig-severity}
\end{figure}

The second heatmap, Figure \ref{fig-severity} on page \pageref{fig-severity}, presents the average CVSS base severity score for CVEs ``compatible'' with a particular CWE--keyword pair. This average score is calculated only for pairs with at least two CVEs; otherwise, the score is set to zero. Relative high scores have been observed for both expected and unexpected combinations of keywords and CWEs. Here are a few examples:

\begin{itemize}
\item \textsl{HMAC} with CWE-798 (hard-coded credentials) and CWE-326 (inadequate encryption strength);
\item \textsl{AES} and \textsl{certificate} with CWE-787 (out-of-bounds write);
\item \textsl{encrypt} with CWE-89 (SQL injection) and CWE-121 (stack-based buffer overflow).
\end{itemize}

\subsection*{Conclusion}

The keyword analysis can be a valuable supplementary tool when working with the NVD dataset, providing insights that might not be immediately apparent through classifications already available in the NVD, such as CWE. A specialized (narrow) set of keywords can be considered in specific scenarios. 

Integrating natural language processing (NLP) tools and techniques can be the next step to enhance the analysis beyond simple keyword search.

\end{document}